# Cybersecurity Challenges in Distributed Control

M.A. Ehsan
School of Engineering and Applied Sciences
University of the District of Columbia
Washington, DC 20008
Email: mdamimul.ehsan@udc.edu

*Abstract*—Cyber-physical systems are becoming core of the most modern systems consisting control, data sharing and real-time monitoring. While centralized control technique has been implemented in the past, recent innovation in distributed control schemes makes it attractive due to various reasons. One of them is the use of state-of-the-art communication protocols that makes the system more robust toward extreme conditions and ensures observability. Thus, as an application of cyber-physical systems, distributed control architectures are prone to various cyber-vulnerability which makes cybersecurity research critical in this application domain. This paper reviews recent researches of distributed control architectures, their cyber-vulnerabilities, and reported mitigation schemes. Finally, some research needs are addressed.

## I. INTRODUCTION

With the advent of technology and proliferation of communication technologies, most of the existing systems are undergoing through a modernising effort such as smart grid being the modified version of power grid etc. The core of these changes is cyber-physical systems and its inter-operable widespread applicability. There has been significant effort in research and development of these technologies in the last decade or more that led to a point where a system maybe designed in the fully integrated form with distributed architectural framework for efficient operation and resource utilization.The scope of this paper is limited to discuss such transformation in the control spectrum, namely distributed control which is being achievable by utilizing such integrated approach of cyber-physical and communication systems. Traditionally, control systems used to be designed as a centralized system, an example being the electric grid control. Even though such large-scale systems consist of multiple layers of operation such as- power generation, transmission and distribution, these three operating domains could be modified for more robust behavior in the distributed form, proposed as smart grid. Without going into much details of power systems, this paper intends to address the need for distributed control in large scale systems as well as its associated challenges, one being the cybersecurity aspect of it.

In order to provide an example of how centralized control of legacy power systems differs from the envisioned fully distributed smart grid, Fig. 1 shows the conceptual model of fully interconnected grid [1].

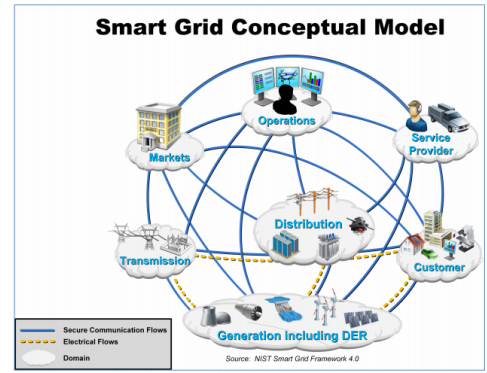

Fig. 1. Updated NIST smart grid conceptual model as described in the NIST smart grid framework 4.0

Due to the high penetration of distributed energy resources (DERs) in the smart grid, the centralized control poses significant challenges due to the following reasons: computational burden due to the high density of controllable resources, communication needs due to wide geographical span, dynamic topology changes, management of high volume of data, and overall reliability and security [2]. In distributed control framework, control task is assigned to various connected distributed units based on the resource availability and urgency. This control hierarchy constitutes primary, secondary and tertiary levels of resources [3], [4].

The implementation of such interconnected systems require extensive use of cyber-physical resources. Therefore, the overall operation is critically dependent on the secure and reliable communication between various sensors, data processors, controllers etc. To ensure, seamless operation, a few industry standard communication protocols developed in the recent years and accepted widely in the technical community are– distributed network protocol 3 (DNP3), Modbus, transmission control protocol/internet protocol (TCP/IP), XML, IEC 61850, and controller area network (CAN) bus [5]–[7]. One major existing lacking of most of these protocols is the lack of encryption due to resource and timing constraints. Therefore, these resources are vulnerable to cyberattacks of various levels and types. A few such example attacks may include but not limited to are denial-of-serviece (DOS), false data injection (FDI), and replay attack [8]–[10]. While the technical challenges of these attacks are critical, the economic and social impacts could be catastrophic [11]. Therefore, there is an eminent research need to fully understand cyber-vulnerabilities and ways to mitigate undesired situation of systems integrated with distributed control architecture.

This paper is organized as follows- Section II provides an overview of distributed control and Section III lists the cyber-security aspect of distributed control and some example cyber-attacks. A few existing challenges and research needs are addressed in Section IV, followed by concluding discussion in Section V.

## II. DISTRIBUTED CONTROL OVERVIEW

Traditionally, centralized control is essentially system-wide while in decentralized control, the interaction between underlying subsystems is considered almost non-existent [2], [12]. Decentralized control has been widely applicable for larger systems such as power grid, however it does not require communication between DERs and is based on droop characteristics [13], [14]. While it may seem to work fine but it becomes challenging to manage high penetration small-scale solar PV and wind generators due to their high degree of variability and low inertia [15]. Droop based control also becomes problematic due to its poor transient performance, not accounting for load dynamics, lack of black start support, distribution network control performance, inefficiency in economic dispatching of DERs, and not knowing how to manage high frequency switching [16]–[18]. Therefore, distributed control is proposed in resolving most of these challenges in a fully integrated power grid also known as smart grid. In distributed control, the grid is divided into multiple subsystems each of which includes DERs, loads and power lines. While this strategy may preserve the autonomy of each operator, it also helps to overcome computational resource constraints which may lead to help resolving various timing limitations [2].

Now, in distributed control, an underlying optimization problem is solved in a distributed manner while ensuring the communication between various controllable assets. Different distributed control techniques are described in the following subsections.

### A. Distributed model predictive control

Model-predictive control (MPC) is an optimization-based control method [19]. It uses the mathematical model of a system for predicting the system behavior. Primarily, it has an objective function that represents the intended system behavior, formalization of operational constraints that needs to be satisfied, system states (measured) at each time step, and uncertainty information of the noise. The key benefits of MPC includes- capability to handle multivariate control problems, easy tuning, and constraint considerations. Fig 2 shows a discrete-time MPC technique where control signal is computed by minimizing the cost function over a finite number of time steps of the system model. In each time step, it calculates set point while minimizing the weighted sum of errors over the intended time window when only the first input is applied and the remaining are updated later on. Finally, this procedure is repeated in a sequence.

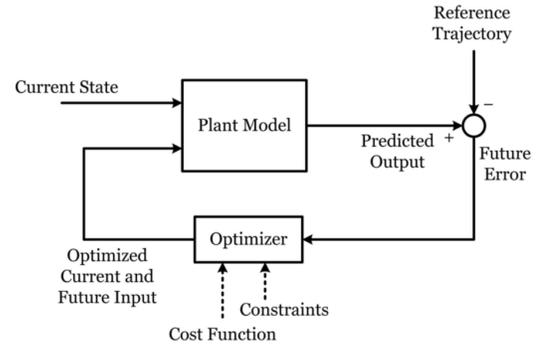

Fig. 2. Model predictive control

### B. Consensus-based control

In the consensus-based approach, distributed optimization problem is solved while offering a flexible formulation ensuring extendability and scalability [20]. An application of this method is illustrated in different DERs converging to a single value. This method achieves global optimality by limited or time-varying communication between nearby instances. Researches show unconstrained consensus-based such algorithms in distributed environment in [20], [21]. The method is then extended to apply in constrained case as well [22].

### C. Agent-based control

The agent-based control is analytically similar to a multi-agent software framework where a problem is decomposed into multiple autonomous components which may interact in a flexible manner to attain a set of objectives, key abstraction model, and explicit technique to describe and manage the complex interplay between agents [23]. This approach is described in Fig 3. In a system level control, it helps resolving drawbacks of hierarchical control by distributing the overall control action into various agents [24].

### D. Decomposition-based control

This method is essentially useful when part of the system becomes islanded or disconnected from the rest [25]. It decomposes the stability analysis process considering input-to-state stability properties of the isolated subsystems and the network topology. It is advantageous due to its flexible operating conditions. A few examples of such methods include auxiliary problem principle (APP), predictor-corrector proximal multiplier method (PCPM), and alternating direction method (ADM).

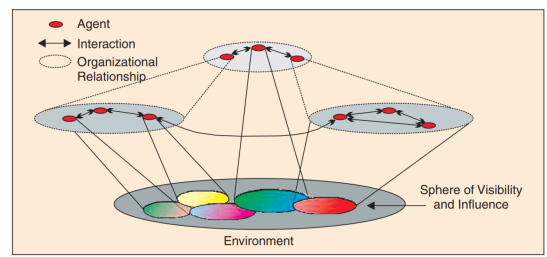

Fig. 3. Agent-based control system (canonical view)

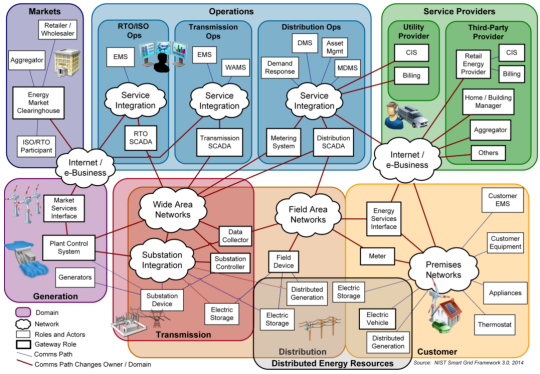

Fig. 4. Legacy communication pathways scenario

## III. Cybersecurity issues in distributed control

This section discusses various cyber vulnerable areas of distributed control architecture and example attacks from literature.

### A. Cyber-vulnerability of distributed control

Prior to discussing the cyber-vulnerable areas of distributed control application, it is important to examine a similar practical system such as power grid. Fig 4 portrays the communication links between various domains and stakeholders of the legacy electric grid. It is a conceptual model as outlined in [1]. It also shows information flow and communication between different subsystems. However, this model does not include the high penetration DER scenarios. At present, the envisioned smart grid includes an extensive provision for various types of high DER integration which is depicted in Fig 5. Therefore, implementation of distributed control strategies to this envisioned power grid is simultaneously critical and challenging. The discussion hence shall focus only around the cyber-vulnerable topic as overall operation of such a complex and large system may become multiverse.

As the need for such discussion is established, now the paper concentrates on the general cyber-vulnerability assessment of distributed control architecture. It includes the risk analysis of a system in a scenario where some communication link is compromised. It may be in virtual or physical or in combination. Physical means permanent damage of a node within the system. Overall, this type of analysis consists of evaluating the impact of an attack on the system dynamics. For the example application case, smart grid system dynamics may include parameters such as voltage, current, frequency, real and reactive power. A criteria, named as 'vulnerability index' of communication links is proposed to examine in [10] based on the oscillation characteristics on the system parameters. The index is distributed over a scale of 1–5 by performing normalization. Net vulnerability index is computed by aggregating all the link level index values.

### B. Cyber-attacks reported in literature

A few common types of cyber-attacks on a distributed control systems are denial-of-service (DOS), false data injection (FDI), and replay attack. While these are only a few known examples from the literature, ideally, the system shall be able to withstand any novel attack as well to ensure a robust characteristics.

*1) DOS attack:* In this cyber attack strategy, the attacker takes control of one or more nodes in the system and keeps the node busy so it cannot communicate with the rest of the system [26]. Therefore it results a transmitting node (or a receiving node) within the network going unresponsive. While DOS attacks has been known for a while, distributed DOS (DDOS) is a modified and more critical form [27]. It consists of three phases and four components (attacker, multiple control primary or handlers, multiple secondary, agents or zombies and a victim or a target machine) as shown in Fig 6. In the first stage, the attacker compromises or infects the handlers which are then use to identify and infect secondary agents that then turns into an army of dummy attackers and works as the attack base to the victim nodes.

*2) FDI attack:* In this attack mode, the attacker interfere with the data packets that are sent and received within the system by manipulating or altering the data at an intermediate point. It may be done in the similar strategy as eavesdropping and the manipulated data could be just some fabricated or noise values that disturb the controller resulting erroneous control signal and eventually leads to greater complexity [28], [29].

*3) Replay attack:* By replay attack, an attacker sends a particular block of previously captured data packets periodically to the original data stream, thus manipulates the actual data [30], [31]. This attack usually takes place when system operates in steady state. The attacker generally gains control over a sensor or node and collects data which is later used to corrupt actual data.

Overall, above listed are a few attacks addressed commonly in the literature [8], [32]–[34]. However, 'Stuxnet'-like [35] attacks that are specifically designed and highly sophisticated may need more attention in the research community to guarantee system resiliency.

## IV. Steps toward ensuring cybersecurity in distributed control

While ensuring a cyber-secured distributed control systems (such as- smart grid architecture) may continue being a priority research area, a few steps toward achieving that shall be discussed in this section.

### A. Communication protocols

Implementation of desired distributed control architecture requires an establishment of reliable communication links between various infrastructures, stakeholders and domains. A similar mapping is shown in Fig 5 for smart grid application case. It requires a set of interoperable communication protocols as various hardware may have different communication requirements. However, the critical matter to ensure is the secure operation of such interoperable communication framework. While the newly deployed devices may have flexibility to select which communication protocol to implement from the design phase, most systems consist of different legacy devices which may not be as flexible and are prone to cyber attacks.

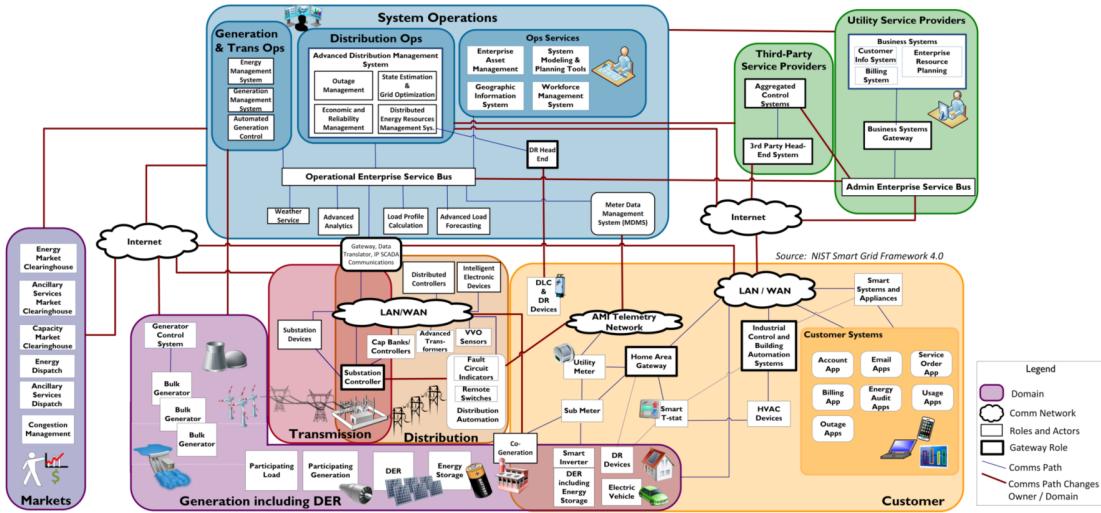

Fig. 5. High-DER communication pathways scenario

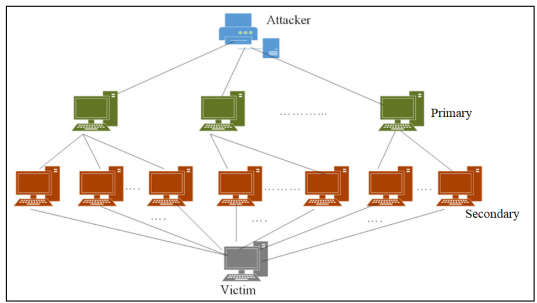

Fig. 6. Structure of a DDOS attack

[1] lists the required communication protocols and their functionalities in the smart grid application. It includes Modbus (does not provide cybersecurity function), ISO/IEC 15118 (EV charging), IEEE 2030.5/SEP2 (DER communication), IEC 61850 (for low latency communication), OCPP, IEC 61968/70 (common information model), and IEEE 1815 (DNP3). All these protocols while are required for ensuring the necessary communication link, in the most cases, they do not provide the protection against cyber attacks. Only IEC 62351-7 (to notify possible attacks) provides some level of cyber vulnerability solutions for IEC 61850 protocol. As of now, most standards and frameworks require custom/organizational cybersecurity and reporting needs to ensure the comprehensive cybersecurity functions described in [36]. Therefore, there is a significant need in developing cyber-secured protocols for reliable operations of systems incorporating distributed control.

### B. High resolution observability

While attacks are eminent and mostly unpredictable, as a system designer, one approach could be to ensure high level of observability across the system. If multiple means of reporting validation can be ensured, even though a communication link/node maybe compromised, the system shall easily identify the attack and take recovery steps with confidence.

### C. Attack detection and mitigation

Technologies in attack detection and mitigation are still in incubation. Literature shows attempts taken to apply various approaches including deep learning and sophisticated data-driven artificial algorithms for this purpose. In addition, to the detection, innovative engineering solution is still needed to design high resilient distributed control systems against adversaries.

### D. Modeling and preparedness

To achieve high degree of reliability, system modeling and various scenario studies are very critical. It not only helps to understand various dynamics under different operating conditions, essentially under-attack situation, it may help toward innovative theoretical break-though as well.

## V. CONCLUSION

This survey paper reviewed existing literature to report associated cyber-risks of distributed control system and reviewed various architectural concepts. Various attack scenario were discussed, essentially for an example case of smart grid or similar application. Finally, a few research needs are identified to ensure the cybersecurity framework functions, namely-identification, protection, detection, response and recovery.

## REFERENCES

[1] A. Gopstein, C. Nguyen, C. O'Fallon, D. Wollman, and N. Hasting, "Nist framework and roadmap for smart grid interoperability standards, release 4.0," 2020.
[2] M. Yazdanian and A. Mehrizi-Sani, "Distributed control techniques in microgrids," *IEEE Transactions on Smart Grid*, vol. 5, no. 6, pp. 2901–2909, 2014.
[3] F.-L. Lian, J. Moyne, and D. Tilbury, "Network design consideration for distributed control systems," *IEEE Transactions on Control Systems Technology*, vol. 10, no. 2, pp. 297–307, 2002.
[4] K. Stouffer, J. Falco, and K. Scarfone, "Guide to industrial control systems (ics) security," *NIST special publication*, vol. 800, no. 82, pp. 16–16, 2011.
[5] V. C. Gungor and F. C. Lambert, "A survey on communication networks for electric system automation," *Computer Networks*, vol. 50, no. 7, pp. 877–897, 2006.

[6] A. Alfergani, A. Khalil, Z. Rajab, M. Zuheir, A. Asheibi, S. Khan, E. H. E. Aboadla, K. A. B. Azna, and M. Tohtayong, "Control of master-slave microgrid based on can bus," in *2017 IEEE Jordan Conference on Applied Electrical Engineering and Computing Technologies (AEECT)*. IEEE, 2017, pp. 1–6.

[7] A. Mehrizi-Sani and R. Iravani, "Potential-function based control of a microgrid in islanded and grid-connected modes," *IEEE Transactions on Power Systems*, vol. 25, no. 4, pp. 1883–1891, 2010.

[8] S. Liu, P. X. Liu, and X. Wang, "Effects of cyber attacks on islanded microgrid frequency control," in *2016 IEEE 20th International Conference on Computer Supported Cooperative Work in Design (CSCWD)*. IEEE, 2016, pp. 461–464.

[9] Z. He, "Reliability evaluation of microgrid cps considering cyber-control-frequency," in *2019 IEEE 2nd International Conference on Electronics Technology (ICET)*. IEEE, 2019, pp. 166–170.

[10] S. Rath, D. Pal, P. S. Sharma, and B. K. Panigrahi, "A cyber-secure distributed control architecture for autonomous ac microgrid," *IEEE Systems Journal*, 2020.

[11] M. Noguchi and H. Ueda, "An analysis of the actual status of recent cyberattacks on critical infrastructures," *NEC Technical Journal, Special Issue Cybersecurity*, vol. 12, no. 2, pp. 19–24, 2019.

[12] M. Ilic and S. Liu, *Hierarchical power systems control: its value in a changing industry*. Springer Science & Business Media, 2012.

[13] A. H. Etemadi, E. J. Davison, and R. Iravani, "A decentralized robust control strategy for multi-der microgrids—part i: Fundamental concepts," *IEEE Transactions on Power Delivery*, vol. 27, no. 4, pp. 1843–1853, 2012.

[14] Y. A.-R. I. Mohamed and E. F. El-Saadany, "Adaptive decentralized droop controller to preserve power sharing stability of paralleled inverters in distributed generation microgrids," *IEEE Transactions on Power Electronics*, vol. 23, no. 6, pp. 2806–2816, 2008.

[15] C. L. DeMarco, C. A. Baone, Y. Han, and B. Lesieutre, "Primary and secondary control for high penetration renewables," *The Future Grid to Enable Sustainable Energy Systems*, 2012.

[16] X. Yu, A. M. Khambadkone, H. Wang, and S. T. S. Terence, "Control of parallel-connected power converters for low-voltage microgrid—part i: A hybrid control architecture," *IEEE Transactions on Power Electronics*, vol. 25, no. 12, pp. 2962–2970, 2010.

[17] M. B. Delghavi and A. Yazdani, "An adaptive feedforward compensation for stability enhancement in droop-controlled inverter-based microgrids," *IEEE Transactions on Power Delivery*, vol. 26, no. 3, pp. 1764–1773, 2011.

[18] J. M. Guerrero, J. C. Vasquez, J. Matas, L. G. De Vicuña, and M. Castilla, "Hierarchical control of droop-controlled ac and dc microgrids—a general approach toward standardization," *IEEE Transactions on industrial electronics*, vol. 58, no. 1, pp. 158–172, 2010.

[19] R. R. Negenborn and J. M. Maestre, "Distributed model predictive control: An overview and roadmap of future research opportunities," *IEEE Control Systems Magazine*, vol. 34, no. 4, pp. 87–97, 2014.

[20] R. Olfati-Saber, J. A. Fax, and R. M. Murray, "Consensus and cooperation in networked multi-agent systems," *Proceedings of the IEEE*, vol. 95, no. 1, pp. 215–233, 2007.

[21] T. Keviczky, F. Borrelli, and G. J. Balas, "Decentralized receding horizon control for large scale dynamically decoupled systems," *Automatica*, vol. 42, no. 12, pp. 2105–2115, 2006.

[22] A. Nedic, A. Ozdaglar, and P. A. Parrilo, "Constrained consensus and optimization in multi-agent networks," *IEEE Transactions on Automatic Control*, vol. 55, no. 4, pp. 922–938, 2010.

[23] N. R. Jennings and S. Bussmann, "Agent-based control systems," *IEEE control systems*, vol. 23, no. 3, pp. 61–74, 2003.

[24] A. L. Dimeas and N. D. Hatziargyriou, "Agent based control of virtual power plants," in *2007 International Conference on Intelligent Systems Applications to Power Systems*. IEEE, 2007, pp. 1–6.

[25] B. Qin, J. Ma, W. Li, T. Ding, H. Sun, and A. Y. Zomaya, "Decomposition-based stability analysis for isolated power systems with reduced conservativeness," *IEEE Transactions on Automation Science and Engineering*, vol. 17, no. 3, pp. 1623–1632, 2020.

[26] G. Carl, G. Kesidis, R. R. Brooks, and S. Rai, "Denial-of-service attack-detection techniques," *IEEE Internet computing*, vol. 10, no. 1, pp. 82–89, 2006.

[27] T. Mahjabin, Y. Xiao, G. Sun, and W. Jiang, "A survey of distributed denial-of-service attack, prevention, and mitigation techniques," *International Journal of Distributed Sensor Networks*, vol. 13, no. 12, p. 1550147717741463, 2017.

[28] Y. Mo and B. Sinopoli, "False data injection attacks in control systems," in *Preprints of the 1st workshop on Secure Control Systems*, 2010, pp. 1–6.

[29] Y. Liu, P. Ning, and M. K. Reiter, "False data injection attacks against state estimation in electric power grids," *ACM Transactions on Information and System Security (TISSEC)*, vol. 14, no. 1, pp. 1–33, 2011.

[30] Y. Mo and B. Sinopoli, "Secure control against replay attacks," in *2009 47th annual Allerton conference on communication, control, and computing (Allerton)*. IEEE, 2009, pp. 911–918.

[31] T. Aura, "Strategies against replay attacks," in *Proceedings 10th Computer Security Foundations Workshop*. IEEE, 1997, pp. 59–68.

[32] J. Liu, Y. Du, S.-i. Yim, X. Lu, B. Chen, and F. Qiu, "Steady-state analysis of microgrid distributed control under denial of service attacks," *IEEE Journal of Emerging and Selected Topics in Power Electronics*, 2020.

[33] B. Zhang, C. Dou, D. Yue, Z. Zhang, and T. Zhang, "A cyber-physical cooperative hierarchical control strategy for islanded microgrid facing with random communication failure," *IEEE Systems Journal*, vol. 14, no. 2, pp. 2849–2860, 2020.

[34] D. B. Rawat and C. Bajracharya, "Detection of false data injection attacks in smart grid communication systems," *IEEE Signal Processing Letters*, vol. 22, no. 10, pp. 1652–1656, 2015.

[35] M. Baezner and P. Robin, "Stuxnet," ETH Zurich, Tech. Rep., 2017.

[36] M. P. Barrett, "Framework for improving critical infrastructure cybersecurity," *National Institute of Standards and Technology, Gaithersburg, MD, USA, Tech. Rep*, 2018.